# Spin wave theory for the triaxial magnetic anisotropy 2D van der Waals antiferromagnet CrSBr

Sergio M. Rezende,[1] Byron Freelon,[2] and Roberto L. Rodríguez-Suárez.[3]

[1]Departamento de Física, Universidade Federal de Pernambuco, 50670-901, Recife, Pernambuco, Brazil.

[2]Physics Department and Texas Center for Superconductivity, University of Houston, Houston, Texas 77204, USA.

[3]Facultad de Física, Pontificia Universidad Católica de Chile, Casilla 306, Santiago, Chile.

**Abstract**

The magnetic properties of two-dimensional (2D) materials have been attracting increasing attention in recent years due to their unique behavior and possible applications in new devices. One material of great interest is the 2D van der Waals (vdW) crystal CrSBr, that exhibits antiferromagnetic (AF) order at low temperatures due to an interlayer AF exchange interaction. Here we present a full quantum spin-wave theory for CrSBr considering three intralayer and one interlayer exchange interactions, and triaxial magnetic anisotropy. The fits of the theoretical results to antiferromagnetic resonance (AFMR) measurements and inelastic neutron scattering data yield reliable values for the seven interaction parameters that can be used to calculate other properties of this interesting material.

*Corresponding author: E-mail: sergio.rezende@ufpe.br

## 1. INTRODUCTION

In the last two decades, the tremendous developments in the techniques for fabricating ultrathin films and heterostructures of a large variety of materials with atomically flat surfaces and interfaces have opened new opportunities for scientific discoveries in two-dimensional (2D) materials and for possible application in future generations of optical, electronic, magnetic, and spintronic devices, among others [1-13]. The magnetic properties, in particular, have been intensively explored in 2D materials and heterostructures for over a decade, motivated mainly by the discoveries of various spintronic phenomena such as the spin-to-charge current conversion [3,4,7, 14-24]. Each material property can serve as a tunable parameter for a specific device purpose, so that a material with multiple properties can serve as a platform for multifunctional devices.

A particular class of 2D materials that have been the focus of intense research efforts inspired by the unprecedented properties of graphene comprises the so-called magnetic van der Waals (vdW) compounds [6,7,13,25-51]. The discovery of magnetic ordering in monolayer $CrI_3$ [28-31] led to the identification of other chromium trihalides such as $CrBr_3$ and $CrCl_3$ [27,32,33] as 2D magnets in the monolayer and few-layer limits and triggered research that led to the development of several bulk crystals and heterostructures made of stacks of 2D magnetic layers held together by van der Waals forces showing stability at ambient conditions.

Initially synthesized in its bulk form in the early 1990s [52], only recently CrSBr has emerged as an interesting and challenging 2D magnetic material [53-77]. Unlike several other 2D vdW magnets, CrSBr is an air-stable material that resists in-ambient degradation over several months [62], can be easily exfoliated [64], and is highly deformable [39]. The investigation of this transition metal chalcohalide has been motivated by its interesting magnetic, electronic, and optical properties with 2D features. CrSBr is a magnetic semiconductor that at room temperature is in the paramagnetic phase [54,65,66]. As the temperature is decreased, a rectangular a-b planar ferromagnetic (FM) phase transition occurs at a Curie temperature of 146 K, and below the Neel temperature ($T_N$) of 132 K these FM layers couple antiferromagnetically along the stacking direction in an A-type antiferromagnetic (AF) ordering along the c-axis [54,55]. In this low-temperature phase, the material exhibits layer-dependent magnetism as well as a tuning dependence based on external stress stimuli [39,56,67,68].

The elastic and magnetic excitations of CrSBr have been investigated by means of several techniques. Inelastic neutron scattering has been used to measure in detail the dispersion relations for magnons propagating in various directions in the FM plane [57]. Microwave driven antiferromagnetic resonance (AFMR) was employed to detect the two zone-center magnons with varying applied magnetic field in the AF phase in a broad temperature range [58]. Time and spatially resolved magneto-optical Kerr effect (MOKE) microscopy showed that two transient strain fields can launch coherent wave packets of magnons at the same frequencies of the AFMR [68]. Ultrafast electron diffraction was employed to investigate the atomic lattice revealing a time-dependence similar to the magnon frequencies [69,77].

In order to understand the range of experimental measurements of the AF magnetic behavior in CrSBr, we have developed a quantum full spin-wave theory for this challenging material considering the three FM intralayer and the interlayer exchange interactions, as well as triaxial magnetic anisotropy. The findings presented here are used to fit the antiferromagnetic resonance (AFMR) measurements [68] and the inelastic neutron scattering data [57] to obtain reliable values for the seven interaction parameters.

## II- QUANTUM THEORY OF SPIN WAVES IN CrSBr

In this section we present a quantum formulation of spin waves for 2D van der Waals magnetic materials aiming to compare with experimental data for CrSBr. Bulk CrSBr crystallizes in the orthorhombic *Pmmm* space group with lattice parameters $a$ = 3.50 Å, $b$ = 4.76 Å, and $c$ = 7.96 Å, exhibiting a vdW layered structure. In each monolayer, the $Cr^{2+}$ spins are in {001} planes, aligned along <100> directions, forming a 2D lattice with rectangular arrangement, with three relevant ferromagnetic exchange interactions. At temperatures below $T_N$ = 132 K, two neighboring layers have spins in opposite directions due to an interlayer exchange interaction, characterizing a bulk A-type

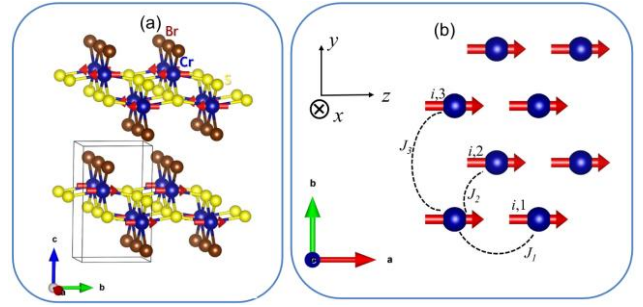

**Figure 1- (Color online)** (a) Illustration of the crystal structure of CrSBr showing the orientations of the $Cr^{2+}$ spins in two neighboring atomic planes in the antiferromagnetic phase. (b) Top view of the projections of the chromium atoms in a single crystal a-b plane showing the three main magnetic exchange interactions.

antiferromagnetic (AF) arrangement, as illustrated in Fig. 1(a).

In order to write the magnetic Hamiltonian, we consider that in a monolayer the spin $i$ interacts via FM exchange parameters $J_1$, $J_2$, and $J_3$ with the neighbor spins $i1$, $i2$, and $i3$ in the same monolayer, as illustrated in Fig. 1(b), and via an AF interlayer exchange $J_I$ with the nearest neighbor spins $j$ and $j'$ in the next monolayers [51,60]. We use a Cartesian coordinate system for the spins projections, with $z$ along the equilibrium direction of spins $S_i$, which is the easy axis in the a-direction, $y$ along the b-direction and $x$ normal to the a-b plane. We also consider an external magnetic field applied in the monolayer plane and three magnetic anisotropies, easy-axis and intermediate along, respectively, the a- and b-directions, and hard-axis in the c-direction [51, 60].

Thus, the Hamiltonian with Zeeman, exchange, and anisotropy interactions has the form [78-83]

$$H = -\sum_{i,j} \gamma \hbar \vec{S}_{i,j} \cdot \vec{H}_0 + \sum_{i,j} -J_1 \vec{S}_{i,j} \cdot \vec{S}_{i1,j1} - J_2 \vec{S}_{i,j} \cdot \vec{S}_{i2,j2} - J_3 \vec{S}_{i,j} \cdot \vec{S}_{i3,j3} + \sum_{i,j} J_I \vec{S}_{i,j} \cdot \vec{S}_{j,i} + \sum D_c (S^x_{i,j})^2 - D_b (S^y_{i,j})^2 - D_a (S^z_{i,j})^2 \quad (1)$$

where $\gamma = g\mu_B/\hbar$ is the gyromagnetic ratio, $g$ is the spectroscopic splitting factor, $\mu_B$ the Bohr magneton, $\hbar$ the reduced Plank constant, $\vec{S}_i$ denotes the spin (in units of $\hbar$) at a generic lattice site $i$, $\vec{H}_0$ is the applied magnetic field, the $J$s are the exchange constants already defined, and $D_a$, $D_b$, and $D_c$ are the three magnetic anisotropy parameters.

We consider the spin Hamiltonian in Eq. (1) and treat the quantized excitations of the magnetic system with the approach of Holstein-Primakoff (HP) [81-84], which consists of transformations that express the spin operators in terms of boson operators that create or annihilate the quanta of spin waves, called magnons. Initially, we replace in the Hamiltonian (1) the Cartesian components of the spin operators $\vec{S}_{i,j}$ in terms of the raising and lowering spin operators $S^{\pm}_{i,j} = S^x_{i,j} \pm i S^y_{i,j}$ [81]. The first HP transformation consists in expressing the spin operators in terms of creation and annihilation operators of spin deviations at the two sublattices, which in the linear approximation become [84]

$$S^+_i = (2S)^{1/2} a_i, \quad S^-_i = (2S)^{1/2} a^\dagger_i, \quad S^z_i = S - a^\dagger_i a_i, \quad (2a)$$

$$S^+_j = (2S)^{1/2} b^\dagger_j, \quad S^-_j = (2S)^{1/2} b_j, \quad S^z_j = -S + b^\dagger_j b_j, \quad (2b)$$

where $a^\dagger_i$, $a_i$, and $b^\dagger_j$, $b_j$, are the creation, destruction operators for spin deviations at sites $i$ and $j$ in opposite spin sublattices, with boson commutation rules $[a_i, a^\dagger_{i'}] = \delta_{ii'}$, $[a_i, a_{i'}] = 0$, $[b_j, b^\dagger_{j'}] = \delta_{jj'}$ and $[b_j, b_{j'}] = 0$. Considering the magnetic field applied along the $a$-axis, with the transformations (2) the Hamiltonian (1) becomes

$$H = \sum_i \gamma\hbar H_0 a^\dagger_i a_i + \sum_j \gamma\hbar H_0 b^\dagger_j b_j$$
$$-\sum_i J_1 S(a^\dagger_i a_{i1} + a^\dagger_i a_{i1} - a^\dagger_i a_i - a^\dagger_{i1} a_{i1}) + J_2 S(a^\dagger_{i2} a_i + a^\dagger_i a_{i2} - a^\dagger_i a_i - a^\dagger_{i2} a_{i2})$$
$$+ J_3 S(a^\dagger_{i3} a_i + a^\dagger_i a_{i3} - a^\dagger_i a_i - a^\dagger_{i3} a_{i3})$$
$$-\sum_j J_1 S(b^\dagger_j b_{j1} + b^\dagger_j b_{j1} - b^\dagger_j b_j - b^\dagger_{j1} b_{j1}) + J_2 S(b^\dagger_{j2} b_j + b^\dagger_j b_{j2} - b^\dagger_j b_j - b^\dagger_{j2} b_{j2})$$
$$+ J_3 S(b^\dagger_{j3} b_j + b^\dagger_j b_{j3} - b^\dagger_j b_j - b^\dagger_{j3} b_{j3})$$
$$+\sum J_I S(a_i b_{ji} + a^\dagger_i b^\dagger_{ji} + b^\dagger_j b_{ji} + a^\dagger_i a_i)$$
$$+\sum_i \frac{D_c S}{2}(a_i a_i + a^\dagger_i a^\dagger_i + a_i a^\dagger_i + a^\dagger_i a_i) - \frac{D_b S}{2}(-a_i a_i - a^\dagger_i a^\dagger_i + a_i a^\dagger_i + a^\dagger_i a_i) + 2D_a S a^\dagger_i a_i$$
$$+\sum_j \frac{D_c S}{2}(b_j b_j + b^\dagger_j b^\dagger_j + b_j b^\dagger_j + b^\dagger_j b_j) - \frac{D_b S}{2}(-b_j b_j - b^\dagger_j b^\dagger_j + b_j b^\dagger_j + b^\dagger_j b_j) + 2D_a S b^\dagger_j b_j$$

(3)

Next, we use the Fourier transforms to express the localized field operators in terms of the collective boson operators that satisfy the commutation rules $[a_k, a^\dagger_{k'}] = \delta_{kk'}$, $[a_k, a_{k'}] = 0$, $[b_k, b^\dagger_{k'}] = \delta_{kk'}$, $[b_k, b_{k'}] = 0$,

$$a_i = N^{-1/2} \sum_k e^{i\vec{k}\cdot\vec{r}_i} a_k, \quad b_j = N^{-1/2} \sum_k e^{i\vec{k}\cdot\vec{r}_j} b_k, \quad (4)$$

where $N$ is the number of spins in each sublattice and $\vec{k}$ is a wave vector, and we have the orthonormality condition

$$N^{-1} \sum_i e^{i(\vec{k}-\vec{k'})\cdot\vec{r}_i} = \delta_{k,k'}. \quad (5)$$

Introducing in Eq. (3) the transformations (4) we obtain for the Hamiltonian written in normal order and without the constants

$$H = \hbar \sum_k (A_k + \gamma H_0) a^\dagger_k a_k + (A_k - \gamma H_0) b^\dagger_k b_k + B_k(a_k b_{-k} + a^\dagger_k b^\dagger_{-k}) + \frac{1}{2} C_k(a_k a_{-k} + b_k b_{-k} + H.c.)$$

(6)

where the new coefficients are given by

$$A_k = \gamma[H_{E1}(1-\gamma_{1k}) + H_{E2}(1-\gamma_{2k}) + H_{E3}(1-\gamma_{3k}) + H_{EI} + H_c/2 - H_b/2 + H_a], \quad (7a)$$

$$B_k = \gamma \gamma_{Ik} H_{EI}, \quad C_k = \gamma(H_c + H_b)/2. \quad (7b)$$

The effective fields in these equations are defined by

$$H_{E1} = 2SJ_1 z_1/g\mu_B, \quad H_{E2} = 2SJ_2 z_2/g\mu_B, \quad H_{E3} = 2SJ_3 z_3/g\mu_B, \quad (8a)$$

$$H_{EI} = SJ_I z_I/g\mu_B, \quad (8b)$$

$$H_c = 2SD_c/g\mu_B, \quad H_b = 2SD_b/g\mu_B, \quad H_a = 2SD_a/g\mu_B. \quad (8c)$$

and the structure factors, given by $\gamma_k = (1/z) \sum_\delta \exp(i\vec{k}\cdot\vec{\delta})$, are

$$\gamma_{1k} = \cos(2k_x a), \quad \gamma_{2k} = \cos(k_x a)\cos(k_y b), \quad (9a)$$

$$\gamma_{3k} = \cos(2k_y b), \quad \gamma_{Ik} = \cos(k_z c) \quad . \quad (9b)$$

The next step consists of performing canonical transformations from the collective boson operators $a^\dagger_k, a_k, b^\dagger_k, b_k$ into magnon creation and annihilation operators $\alpha^\dagger_k, \alpha_k, \beta^\dagger_k, \beta_k$. This is done by means of the Bogoliubov transformation [81-83]

$$a_k = u_k \alpha_k - v_k \beta^\dagger_{-k}, \quad (10a)$$

$$b^\dagger_{-k} = -v_k \alpha_k + u_k \beta^\dagger_{-k}. \quad (10b)$$

Substituting these expressions in Eq. (6) and imposing that the Hamiltonian be cast in the diagonal form

$$H = \sum_k \hbar(\omega_{\alpha k} \alpha^\dagger_k \alpha_k + \omega_{\beta k} \beta^\dagger_k \beta_k), \quad (11)$$

where $\omega_{\alpha_k}$ and $\omega_{\beta_k}$ are the frequencies of the two magnon modes, one can find the frequencies and the transformation coefficients. The frequencies are given by [82,83]

$$\omega^2_{\alpha,\beta} = A^2_k + (\gamma H_0)^2 - (C^2_k + B^2_k) \pm 2[C^2_k B^2_k + \gamma^2 H^2_0 (A^2_k - B^2_k)]^{1/2}. \quad (12)$$

This result will be used in the next section to fit the experimental data for magnons in CrSBr.

### III. APPLICATION TO ANTIFERROMAGNETIC RESONANCE AND INELASTIC NEUTRON SCATTERING EXPERIMENTS

The antiferromagnetic resonance (AFMR) experiment consists of driving the $k = 0$ magnons with *rf* microwave radiation when the sample is in an applied static magnetic field. Under these conditions, the *rf* absorption is observed when the frequency and field match the resonance equation. AFMR experiments with a scanning magnetic field intensity were carried in bulk crystals of CrSBr in a temperature range covering the whole AF phase by Cham et al. [58]. In order to compare our theory with the AFMR data, we use Eq. (12) with the coefficients defined in Eqs. (7a) and (7b) with the structure factors in Eqs. (9a) and (9b) calculated for $k = 0$, that is, $\gamma_{1k} = \gamma_{2k} = \gamma_{2k} = \gamma_{Ik} = 1$. In this case one can show Eq. (12) gives for the two magnon frequencies

$$\omega_\alpha = \gamma \{H_0^2 + H_{EI}(2H_a + H_c - H_b) + H_a(H_a + H_c - H_b) - H_b H_c \\ + 2[H_{EI}^2(H_c + H_b)^2/4 + H_0^2 H_{EI}(2H_a + H_c - H_b) + H_0^2(H_a + H_c/2 - H_b/2)^2]^{1/2}\}^{1/2}, \quad (13)$$

$$\omega_\beta = \gamma \{H_0^2 + H_{EI}(2H_a + H_c - H_b) + H_a(H_a + H_c - H_b) - H_b H_c \\ - 2[H_{EI}^2(H_c + H_b)^2/4 + H_0^2 H_{EI}(2H_a + H_c - H_b) + H_0^2(H_a + H_c/2 - H_b/2)^2]^{1/2}\}^{1/2}. \quad (14)$$

Note that, since the AF exchange interaction acts only between layers, the AFMR frequencies do not depend on the intralayer FM exchange parameters. One can show that for the case of a system with two anisotropies, making $H_a = 0$ and $H_b \to -H_a$, Eqs. (13) and (14) agree with Eqs. (4) and (5) of [58] obtained with a semiclassical dynamic magnetization approach model for two anisotropies in CrSBr. Also, note that for $H_0 = 0$, the two magnon frequencies become

$$\omega_\alpha = \gamma[H_c(2H_{EI} + H_a)]^{1/2}, \quad (15)$$

$$\omega_\beta = \gamma[H_a(2H_{EI} + H_c)]^{1/2}. \quad (16)$$

Figure 2 shows the measured field dependence of the AFMR frequencies measured at several temperatures with the magnetic field along the *a*-axis of a single crystal bulk CrSBr [58] and the fits with Eqs. (13) and (14). The values obtained from the fits for the interlayer AF exchange and anisotropy fields are very sensitive to the variation of the frequencies with magnetic field and temperature. Figure 3 shows the temperature dependence of the four effective fields obtained from the fits. The fact that all fields decrease with increasing temperature and tend to vanish above $T_N$ is consistent with the shift to lower frequencies of the two modes and the decrease in the gap between them shown in Fig. 2. Note that the error bars in Fig. 3 result from the fact that the fits are relatively insensitive to variations of the field parameters in the range of the bars. Note also that the fits made with the two-anisotropy model of Ref. [58] result in a value of the hard-axis anisotropy field $H_c$ quite larger than the interlayer exchange field, while the model with three anisotropies used here gives the opposite.

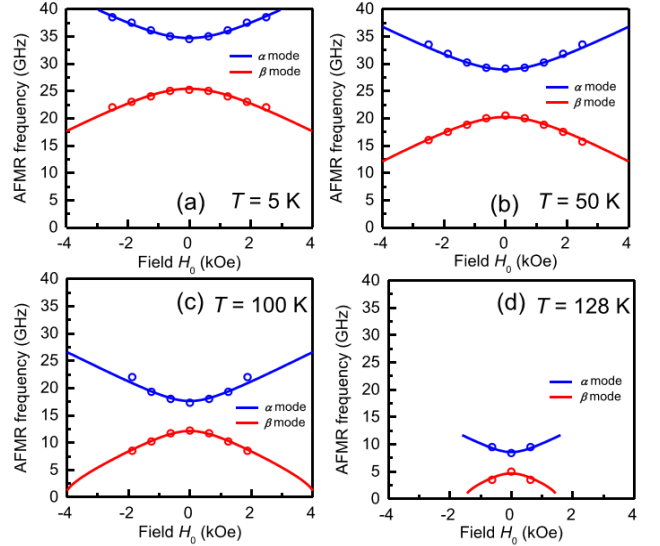

**Figure 2:** (color online) Fits of the theoretical $k = 0$ magnon frequencies (solid lines) to the discretized AFMR data of Cham et al. [58] (symbols) for the 2D vdW AF CrSBr at several temperatures as indicated.

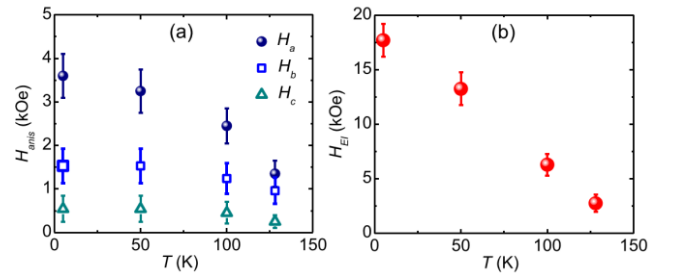

**Figure 3:** (color online) Variation with temperature of the anisotropy (a) and AF exchange (b) fields obtained from the fits of the theoretical frequencies of the tow $k = 0$ magnon frequencies measured in CrSBr [58] as shown in Fig. 2.

Next, we compare the results of the spin-wave theory with the magnon dispersion relations measured by Scheie et al. [57] in a bulk crystal of CrSBr using inelastic neutron scattering with the wave vector in the *a-b* plane, at $T = 5$ K and no applied magnetic field. The symbols in Figs. 4(a) and 4(b) represent the magnon energy as a function of the reduced wave number $q = k/k_m$ ($k_m$ is maximum value of $k$ in each direction) measured for $k$ along the *a*-axis and along the diagonal of the *a-b* axes. The solid lines represent the least square deviation fits obtained with Eq. (12) considering for the AF interlayer exchange and the three anisotropy fields the same values obtained from the fits of Eqs. (13) and (14) to the AFMR data for $T = 5$ K, namely, $H_{EI} = 17.68$ kOe, $H_a = 3.60$ kOe, $H_b = 1.53$ kOe, and $H_c = 0.54$ kOe. The solid curves in Figs. 4(a) and 4(b) represent the fits obtained with the intralayer FM exchange fields $H_{E1} = 448$ kOe, $H_{E2} = 1\,952$ kOe, and $H_{E3} = 864$ kOe. The intralayer exchange parameters obtained from these fields with Eqs. (8a) are $J_1 = 1.73$ meV, $J_2 = 3.77$ meV, and $J_3 = 1.84$ meV. These values are similar but not equal to the values $J_1 = 1.9$ meV, $J_2 = 3.38$ meV, and $J_3 = 1.67$ meV obtained in Ref. [57], and are quite different from the values $J_1 = 3.54$ meV, $J_2 = 3.08$ meV, and $J_3 = 4.15$ meV obtained with first-principle calculations [56].

Note that in the scale of Figs. 4(a) and 4(b), the frequencies of modes $\alpha$ and $\beta$ are indistinguishable because the intralayer FM exchange fields are three orders of magnitude larger than the interlayer AF exchange and anisotropy fields. Figures 4(c) and 4(d) show a zoom near the zone center, where one sees that the frequencies of the $\alpha$ and $\beta$ modes are indeed different for $k < 0.02\ k_m$, as demonstrated in the AFMR experiments. Note that for $k = 0$ the energies of the two modes are 0.105 meV and 0.144 meV, corresponding to frequencies of 25.4 GHz and 34.8 GHz for the AFMR modes at $T = 5$ K in zero field, as in Fig. 2(a).

## IV. CONCLUSIONS

In conclusion, we have developed a full spin-wave theory to calculate the magnon dispersion relations for the 2D van der Waals (vdW) layered crystal of CrSBr in the antiferromagnetic phase. We have considered the three FM intralayer and the AF interlayer exchange interactions, as well as triaxial magnetic anisotropy. The fits of the theoretical results to antiferromagnetic resonance (AFMR) measurements of Cham et al. [58] and to inelastic neutron scattering data of Scheie et al. [57] yield reliable values for the seven interaction parameters that should be useful for further exploring the magnetic properties of this interesting material.

## ACKNOWLEDGEMENTS

The research at Universidade Federal de Pernambuco was supported by Conselho Nacional de Desenvolvimento Científico e Tecnológico (CNPq), Coordenação de Aperfeiçoamento de Pessoal de Nível Superior (CAPES), Financiadora de Estudos e Projetos (FINEP), Fundação de Amparo à Ciência e Tecnologia do Estado de Pernambuco (FACEPE), and INCT of Spintronics and Advanced Magnetic Nanostructures (INCT-SpinNanoMag), CNPq Grant No. 406836/2022-1. BF was partially supported by the U.S. DOE, BES, under Award No. DE-SC0024332, and acknowledges support from the U.S. Air Force Office of Scientific Research and Clarkson Aerospace Corp. under Award FA9550-21-1-0460. The research at PUC/Chile was supported by Fondo Nacional de Desarrollo Científico y Tecnológico (FONDECYT) Grant No. 1130705, and FONDEQUIP projects EQM180103 and EQM190136.

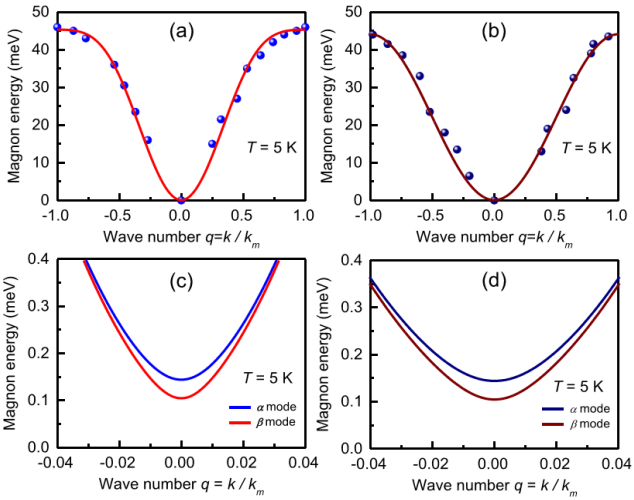

**Figure 4:** (color online) (a) and (b)- Fits of the theoretical magnon dispersion relations (solid lines) for the 2D AF CrSBr to the inelastic neutron scattering data of Scheie et al. [57] (symbols) for the wave vector along the a-axis in (a) and along the diagonal of the a and b axes in (b). (c) and (d)- Zoom of the theoretical magnon dispersion relations at the Brillouin zone center with *k* along the same directions as in the curves above.